\documentclass[
reprint,
superscriptaddress,
showkeys, 
amsmath,amssymb,
aps,pre,
]{revtex4-1}
\usepackage{graphicx}
\usepackage{dcolumn}
\usepackage{bm}

\begin{document}
\preprint{APS/123-QED}

\title{Diversity of dynamical behaviors due to initial conditions: exact results with extended Ott--Antonsen ansatz for identical Kuramoto--Sakaguchi phase oscillators}

\author{Akihisa Ichiki}
\email{ichiki@chem.material.nagoya-u.ac.jp}
\affiliation{Institutes of Innovation for Future Society, Nagoya University, Furo-cho, Chikusa-ku, Nagoya 464-8603, Japan}
\author{Keiji Okumura}
\email{kokumura@ier.hit-u.ac.jp}
\affiliation{Institute of Economic Research, Hitotsubashi University, Kunitachi, Tokyo 186-8603, Japan}
\date{\today}

\begin{abstract}
The Ott--Antonsen ansatz is a powerful tool to extract the behaviors of coupled phase oscillators, but it imposes a strong restriction on the initial condition. 
Herein, an extension of the Ott--Antonsen ansatz is proposed to relax the restriction, 
enabling the systematic approximation of the behavior of a globally coupled phase oscillator system with an arbitrary initial condition. 
The proposed method is validated on the Kuramoto--Sakaguchi model of identical phase oscillators. The method yields cluster and chimera-like solutions that are not obtained by the conventional ansatz. 
\end{abstract}


\maketitle

\section{Introduction}
Coupled oscillator systems are important in both the pure science of synchronization phenomena~\cite{winfree1967biological, dorfler2014synchronization, RevModPhys.77.137} and engineering applications such as electric power~\cite{PhysRevE.93.032222} and wireless communication networks~\cite{Guilera}. It also plays a central role in understanding biological phenomena such as neural networks~\cite{Novikov, Hannaye1701047} and the synchronous rhythm of cardiomyocytes~\cite{hayashi2017community}. The most representative models of coupled phase oscillators are the Kuramoto model~\cite{kuramoto1984chemical} and its generalization, the Kuramoto--Sakaguchi model~\cite{sakaguchi1986soluble}. In recent years, Ott and Antonsen~\cite{ott2009long} proposed a powerful ansatz to analyze these models, and it has deepened the understanding of behaviors of coupled oscillator systems. The Ott--Antonsen ansatz~(OAA) has been successfully employed in systems with distributed natural frequencies~\cite{kawamura2010phase, martens2009exact, hong2012mean} and systems with external driving~\cite{childs2008stability, schwab2012kuramoto}. The OAA essentially reduces a system consisting of numerous coupled phase oscillators to a two-dimensional nonlinear oscillator system~\cite{ott2009long, goldobin2018collective}.  The process is conventionally understood as follows: if the initial distribution of oscillator phases is set to a two-parameter distribution family called the Poisson kernel, the phase distribution after time evolution remains in the Poisson kernel~\cite{marvel2009identical}.  
Thus, the OAA strongly restricts the initial condition, although it is effective for understanding a globally coupled oscillator system.  
In general, the origin of diversity in dynamical behaviors can be rooted in both the native properties of oscillators and the initial condition. The restriction of the OAA hinders the full understanding of the dependence of system behaviors on initial conditions. 
In the present paper, the OAA is extended to systematically relax the restrictions on the initial distribution for understanding the initial-condition dependence of complicated behaviors in phase oscillator systems. 

The present paper makes three main claims. Firstly, the Poisson kernel appearing in the OAA is claimed to be equivalent to a Cauchy--Lorentz distribution (CLD). Therefore, the OAA is interpreted as follows: if the initial phase distribution is set to a CLD, the phase distribution remains in the distribution family of CLD. Secondly, as an extension of the conventional OAA, it is claimed that if the initial phase distribution is set to a superposition of CLDs, the phase distribution remains in the superposition of CLDs. Consequently, an arbitrary initial condition can be analyzed systematically by approximating the phase distribution as a superposition of CLDs.  
Thirdly, compared to the conventional OAA, the extended version is more helpful to understand complicated behaviors of the system. To show the advantage, the extended OAA is employed for the Kuramoto--Sakaguchi model of identical phase oscillators, and it yields a variety of dynamical behaviors including a cluster solution~\cite{golomb1992clustering, gong2019repulsively} that could not be obtained by the conventional OAA. 


\section{Ott--Antonsen ansatz}\label{secOAA}
To overview the derivation of the OAA, let us consider a system of $N$ phase oscillators globally coupled via mean-field couplings, i.e., the Kuramoto--Sakaguchi model~\cite{sakaguchi1986soluble}: 
\begin{eqnarray}
\dot{\theta}_i = \omega_i - \dfrac{K}{N}\displaystyle\sum_{j=1}^N \sin\left(\theta_i - \theta_j + \alpha\right)\,,\label{original_dyn}
\end{eqnarray}
where $\theta_i$ denotes the phase of the $i^{\rm th}$ oscillator, $\omega_i$ the natural frequency of the oscillator $i$, and $K > 0$ the coupling constant. A constant $\alpha$ determines whether the coupling is attractive ($\cos\alpha > 0$) or repulsive ($\cos\alpha < 0$). Many interesting phenomena are known for cases with distributed natural frequencies~\cite{kawamura2010phase, hong2012mean, martens2009exact, nagai2010noise, battogtokh2002coexistence, abrams2004chimera}.
The natural frequency distribution is assumed to be $g(\omega)$. 

The phase distribution of the oscillators is discrete when the number of oscillators is finite but continuous in the thermodynamic limit $N \to\infty$.  Let us consider a situation where the phases of the oscillators are given according to a certain distribution. When the number of oscillators is finite, the empirical (observed) distribution and true distribution for the phases are different. However, as the number of oscillators increases, the empirical distribution approaches the true distribution. Naturally, the true distribution is well approximated by the empirical distribution when the number of oscillators is sufficiently large. In this situation, by the self-averaging property, the empirical distribution of oscillator phases with a natural frequency $\omega$, 
\begin{eqnarray}
P(\theta, t|\omega) := \dfrac{1}{N_\omega}\sum_{j\in \omega}\delta\left(\theta - \theta_j(t)\right),\label{empiricalProb}
\end{eqnarray}
evolves by the following nonlinear Fokker--Planck equation~\cite{kuramoto1984chemical, strogatz2000kuramoto, RevModPhys.77.137}: 
\begin{eqnarray}
&&\dfrac{\partial P(\theta, t|\omega)}{\partial t} = -\dfrac{\partial}{\partial \theta}\left[\omega - KQ(\theta, t)\right]P(\theta, t|\omega)\,,\label{FPE}\\
&&Q(\theta, t) := \dfrac{1}{2i}\left[z^\ast(t) e^{i(\theta + \alpha)} - z(t) e^{-i(\theta+\alpha)}\right]\,,
\end{eqnarray}
where $z^\ast$ indicates the complex conjugate of $z$. 
In Eq.~(\ref{empiricalProb}), $N_\omega$ denotes the number of oscillators with a natural frequency $\omega$, and the sum is taken over all oscillators of a natural frequency $\omega$. 
The Kuramoto order parameter $z(t)$ is defined as~\cite{kuramoto1984chemical, kuramoto1987statistical} 
\begin{eqnarray}
z(t) := \displaystyle\sum_{j=1}^N \exp\left[i\theta_j(t)\right]\,.
\end{eqnarray}
In the thermodynamic limit $N\to\infty$, $z(t)$ is evaluated as 
\begin{eqnarray}
z(t) = \int\,d\omega g(\omega)\int\,d\theta \exp\left(i\theta\right) P(\theta, t|\omega)\,.
\end{eqnarray}
In the study of phase oscillators, the phase is often defined in the domain $0\le \theta <2 \pi$. However, in this paper, the phase is defined in $-\infty <\theta <\infty$. Such an extension of the domain of $\theta$ does not affect the dynamics given by Eq.~(\ref{original_dyn}). 
Therefore, the phase distribution $P(\theta, t | \omega) $ is also defined in the domain $-\infty <\theta <\infty$. The observed phase is $\theta \bmod 2 \pi$, but our definition of phase takes into account how many revolutions the oscillators have made around the origin. Then, the phase distribution $P_{\rm obs}(\theta, t | \omega)$ based on the observed $\theta \bmod 2 \pi$ is given using our phase distribution as $P_{\rm obs}(\theta, t | \omega) = \sum_ {n =-\infty} ^ \infty P(\theta + 2 \pi n, t | \omega)$, where the domain of $P_{\rm obs} (\theta, t | \omega) $ is $ 0 \le \theta <2 \pi$. Note that $P(\theta, t | \omega)$ satisfies the natural boundary condition, i.e., it approaches zero as $\theta\to\pm\infty$. Otherwise, $P_{\rm obs}(\theta, t|\omega)$ diverges according to its relation to $P(\theta, t|\omega)$, which is nonnegative. 

It is convenient to apply the Fourier transform to both sides of Eq.~(\ref{FPE}). Denoting $p_s^\omega (t) := \int_{-\infty}^\infty d\theta\, \exp\left(-is\theta\right) P(\theta, t|\omega)$, the Fokker--Planck equation~(\ref{FPE}) is expressed as 
\begin{eqnarray}
\dot{p}_s^\omega(t) = -is\omega p_s^\omega(t) + \dfrac{sK}{2}\left[z^\ast e^{i\alpha}p_{s-1}^\omega - z e^{-i\alpha}p_{s+1}^\omega\right]\,.\label{dyn}
\end{eqnarray}
To obtain this expression, the natural boundary condition for $P(\theta, t|\omega)$ was used. 
Note that $p_s^\omega(t)$ is related to the characteristic function $\phi_\omega(s, t) := \int\,d\theta\exp\left(is\theta\right)P(\theta, t|\omega)$ for the empirical distribution by $p_s^\omega(t) = \phi_\omega(-s, t)$. 
In general, Eq.~(\ref{dyn}) has an infinite hierarchy and cannot be solved. In order to resolve the hierarchy, Ott and Antonsen~\cite{ott2009long} assumed that there exists an appropriate complex variable  $A_\omega(t)$ and 
\begin{eqnarray}
p_s^\omega(t) = A_\omega^s(t)\label{OAansatz}
\end{eqnarray}
holds for all nonnegative $s \ge 0$. 
As Ott and Antonsen stated in their paper~\cite{ott2009long}, in Kuramoto oscillator systems, both partially synchronized and desynchronized solutions can be expressed in the form of Eq.~(\ref{OAansatz}). The original motivation for this substitution is to restrict the discussion to a distribution family that includes well-known stationary solutions. Note that this substitution restricts the possibly considered distribution family $P(\theta, t|\omega)$. 
This ansatz for $p_s^\omega$ is called the OAA. The OAA is found to be equivalent to 
\begin{eqnarray}
\phi_\omega(-s, t) = \exp\left[-s\gamma_\omega(t) + is\mu_\omega(t)\right]
\end{eqnarray}
for all nonnegative $s \ge 0$ with appropriate real variables $\gamma_\omega(t)$ and $\mu_\omega(t)$.  
The distribution with such a characteristic function is a CLD.   
The parameters $\mu_\omega$ and $\gamma_\omega$ play the roles of the location and half-width at half-maximum of the peak in CLD, respectively. 

The Ott--Antonsen manifold is a two-dimensional manifold that is invariant under a time evolution, and it has been discussed in relation to the Poisson kernel~\cite{marvel2009identical}. Because the Poisson kernel is defined in the domain of $0\le \theta < 2\pi$, from the above argument, it is a superposition of a CLD by shifting $2\pi n$ ($n\in\mathbb{Z}$). CLD is a family of distributions characterized by two parameters: peak location and half-width at half-maximum. The manifold formed by these two parameters is the Ott--Antonsen manifold. In discussions on coupled phase oscillator systems using the OAA, the natural frequency distribution $g (\omega)$ is often assumed to be a CLD~\cite{ott2009long}. 
However, if the initial phase distribution in the ensemble of oscillators with the same natural frequency is taken as a CLD, the phase distribution at any time is given by a CLD, regardless of the shape of the frequency distribution $g(\omega)$. 

By introducing a complex variable $A_\omega := \exp\left[-\gamma_\omega(t) + i\mu_\omega(t)\right]$, from the evolution equation~(\ref{dyn}) with the OAA~(\ref{OAansatz}), the evolution for $A_\omega$ is given as 
\begin{eqnarray}
\dot{A}_\omega = i\omega A_\omega + \dfrac{K}{2}z e^{-i\alpha} - \dfrac{K}{2}z^\ast e^{i\alpha}A_\omega^2\label{dyn_A_ori}
\end{eqnarray}
with $z(t) = \int\,d\omega g(\omega)A_\omega(t)$.  
When a phase distribution is assumed to be a CLD, the evolution equation~(\ref{dyn_A_ori}) is straightforwardly obtained by averaging the dynamics~(\ref{original_dyn}). By introducing a complex variable $A_i = \exp\left[i \theta_i (t) \right]$, Eq.~(\ref{original_dyn}) is rewritten in the same form as Eq.~(\ref{dyn_A_ori}): 
\begin{eqnarray}
\dot{A} _i = i \omega A_i + \dfrac{K}{2} z e^{-i\alpha}-\dfrac{K}{2} z ^ \ast e^{i\alpha} A_i ^ 2\,. \label{1body}
\end{eqnarray}
When the phase distribution for oscillators with the natural frequency $\omega$ is given by a CLD 
\begin{eqnarray}
P_{\rm CL}\left(\theta| \mu_\omega(t), \gamma_\omega(t)\right) := \dfrac{1}{\pi}\dfrac{\gamma_\omega(t)}{\left[\theta - \mu_\omega(t)\right]^2 + \gamma_\omega^2(t)}\,,
\end{eqnarray}
the averaged quantity $\bar{A}_\omega := \sum_{j\in\omega} A_j / N_\omega$ is given by the pole of $P_{\rm CL}\left(\theta|\mu_\omega(t), \gamma_\omega(t)\right)$ in the thermodynamic limit as $\bar{A}_\omega(t) = \exp\left(i\theta\right)|_{\theta = \mu_\omega(t) + i\gamma_\omega(t)} = A_\omega$.  
Thus, the set of oscillators obeying the CLD is reduced to a single oscillator with a complex phase evolving with Eq.~(\ref{1body}). 
It is concluded that the OAA is a reduction method for degrees of freedom owing to the representative property of a pole in a CLD.  

\section{Extension of OAA}\label{secExOAA}
As discussed in section~\ref{secOAA}, the conventional OAA considers only the case where the initial phase distribution is a CLD.  
In the analysis of systems with distributed natural frequencies~\cite{ott2009long, martens2009exact, kawamura2010phase, hong2012mean}, the OAA restricts the initial distribution of a phase oscillator group with each frequency to a CLD.  Consequently, the CLD for each frequency evolves as per Eq.~(\ref {dyn_A_ori}). 

To extend the conventional OAA, it is worth mentioning that, in Eq.~(\ref{dyn_A_ori}), oscillators belonging to different CLDs interact only through the Kuramoto order parameter.  
This is true even when the natural frequency distribution is significantly sharp.  Then, it is possible to consider the case where the width of the natural frequency distribution around a certain frequency $\omega$ approaches zero.  In this case, it is concluded that the phase distribution of oscillators with the natural frequency $\omega$ is given by a superposition of CLDs satisfying the OAA.  
To formulate this idea, for simplicity, let us consider a set of phase oscillators with a single natural frequency. By dividing the $N$ oscillators into $M$ groups, the empirical distribution of the $\nu^{\rm th}$ group is formally defined as $P_\nu(\theta, t) = \frac{1}{N_\nu}\sum_{j\in \Omega_\nu}\delta\left(\theta - \theta_j(t)\right)$, where the sum is taken for all oscillators in the $\nu^{\rm th}$ group $\Omega_\nu$ and $N_\nu$ is the number of oscillators in $\Omega_\nu$ of $\mathcal{O}(N)$.  Consequently, oscillators belonging to different groups interact only via the Kuramoto order parameter.  If the initial distribution of $P_\nu$ is a CLD, the time evolution is exactly given by the OAA.  In other words, if the initial phase distribution is given as a superposition of $M$ CLDs, the empirical distribution at any time remains to be a superposition of $M$ CLDs: 
\begin{eqnarray}
P(\theta, t) = \displaystyle\sum_{\nu=1}^M r_\nu P_{\rm{CL}}(\theta|\mu_\nu(t), \gamma_\nu(t))\,,
\end{eqnarray}
where $r_\nu := N_\nu / N$ is the ratio of the $\nu^{\rm th}$ distribution.  Because the conventional OAA holds for each group, the evolution equation is given as 
\begin{eqnarray}
\dot{A}_\nu &=& i\omega A_\nu + \dfrac{K}{2}z e^{-i\alpha} - \dfrac{K}{2}z^\ast e^{i\alpha}A_\nu^2\,,\label{SKOAA}\\
z(t) &=& \displaystyle\sum_{\nu=1}^M r_\nu A_\nu(t)\,,\label{def_z}
\end{eqnarray}
where $A_\nu$ corresponds to the characteristic function of $P_\nu(\theta, t)$, which is assumed to be a CLD.  
In this framework, the Ott--Antonsen manifold is extended to $2M$-dimensions.  
That is, if the superposition of arbitrary CLDs is set as an initial phase distribution, the time evolution of the phase distribution is exactly given. 
This fact is useful for systematically approximating the behavior of a system starting from an arbitrary initial distribution.  The approximated system behavior can be obtained with arbitrary accuracy if the initial phase distribution is approximated as a superposition of CLDs to the required accuracy.  

Note that $A_\nu = \exp\left(-\gamma_\nu + i \mu_\nu\right)$ plays the role of a local Kuramoto order parameter for the $\nu^{\rm th}$ group because $A_\nu$ corresponds to the ensemble average of $\exp\left(i\theta_j\right)$ in the group as $A_\nu = \frac{1}{N_\nu}\sum_{j\in\Omega_\nu}\exp\left(i\theta_j\right)$. Therefore, $\gamma_\nu =-\ln | A_\nu |$ indicates the degree of desynchronization in the group. $\gamma_\nu = 0$ and $\gamma_\nu\to\infty$ indicate phase-locked and desynchronized solutions, respectively. In the case of $M = N$ and $r_\nu = 1/N$ for all $\nu$, Eq.~(\ref{SKOAA}) reproduces the dynamics for each oscillator Eq.~(\ref{1body}).  

Equation~(\ref{SKOAA}) can also be expressed in terms of $\gamma_\nu $ and $\mu_\nu$ as 
\begin{eqnarray}
\dot{\gamma}_\nu & = & -K e^{-\gamma}\sinh\gamma_\nu\cos\left(\mu-\alpha-\mu_\nu\right)\,, \label{gamma_dyn} \\
\dot{\mu}_\nu & = &\omega + K e^{-\gamma}\cosh\gamma_\nu\sin\left(\mu-\alpha-\mu_\nu \right)\,, \label{mu_dyn}
\end{eqnarray}
where $\gamma$ and $\mu$ are defined by the Kuramoto order parameter as $z = \exp\left(-\gamma + i \mu\right)$. 
Equations~(\ref {gamma_dyn}) and (\ref {mu_dyn}) are useful for directly evaluating the degree of desynchronization in the group but are unsuitable for numerical calculations because the constraint $\gamma_\nu \ge 0$ should be imposed and $\gamma_\nu$ may diverge. It is recommended to solve Eqs.~(\ref{SKOAA}) for stable numerical calculations. 

To guarantee the correctness of the above extension of the OAA, it is worth mentioning its mathematical background. The conventional OAA is interpreted as follows: if the initial phase distribution is a CLD, the phase distribution remains in the CLD family at any arbitrary time. The evolution equation for the distribution function is given through the characteristic function. The characteristic function is the Fourier transform of the distribution function, and the Fourier transform is a linear transformation. Therefore, the evolution of the superposition of CLDs is given by the superposition of characteristic functions, and the OAA holds for each CLD.

\section{Deviation from true solutions}\label{secValid}
Equation~(\ref{SKOAA}) is exact in the thermodynamic limit when the initial phase distribution is given by a superposition of CLDs. However, such a situation seems somewhat unrealistic. In some cases, it is not practical to express the initial phase distribution by a superimposition of CLDs. In other cases, the thermodynamic limit does not hold, because the system consists of a finite number of oscillators. This section discusses how the proposed extended OAA deviates from reality.

Even when the number of oscillators $N$ is finite or the oscillator distribution is not given by a superposition of CLDs, Eq.~(\ref{1body}) for each oscillator exactly holds. Considering the ensemble average of $A_j$ in the $\nu^{\rm th}$ group, i.e., the local Kuramoto order parameter for the $\nu^{\rm th}$ group $A_\nu (t): = \frac{1}{N_\nu}\sum_{j \in\Omega_\nu}A_j (t)$,  the exact dynamics for $A_\nu$ is given as 
\begin{eqnarray}
\dot{A}_\nu = i\omega A_\nu + \dfrac{K}{2} z e^{-i\alpha}-\dfrac{K}{2} z^\ast e^{i\alpha}\left\langle A^2\right\rangle_\nu\,,\label {exact_dyn}
\end{eqnarray}
where $\left\langle A^2\right\rangle_\nu := \frac{1}{N_\nu}\sum_{j \in\Omega_\nu}A_j^2$. This equation differs from Eq.~(\ref{SKOAA}) only in the third term on the right-hand side. Note that $\left\langle A^2\right\rangle_\nu$ becomes equal to $A_\nu^2$ when the oscillator phases in $\Omega_\nu$ obeys a CLD. If the phase distribution is not a CLD, the values of $\{A_j \}_{j\in\Omega_\nu}$ must be known to evaluate $\left\langle A^2\right\rangle_\nu$. 
Here, the deviation between the extended OAA and the realistic system can always be evaluated by the third term on the right-hand side in Eq.~(\ref {exact_dyn}). 
When the number of oscillators is finite, the phase distribution is regarded not as a superposition of CLDs but as a discrete distribution. 
Thus, the deviation from the OAA due to both the finite size effect and non-CLD phase distribution can be evaluated based on the same criterion. 

With the proposed method, in principle, an approximation of the initial distribution as a superposition of $M$ CLDs yields an approximation of the time evolution of a system with arbitrary accuracy. Any initial distribution can be approximated well if $M$ is sufficiently large. $M = N$ reproduces the true dynamics by Eq.~(\ref{1body}), which is equivalent to the original dynamics given by Eq.~(\ref{original_dyn}). However, in practice, calculations with a large $M$ are impossible. At present, there is no clear criterion for determining the appropriate value of $M$ that effectively reduces the deviation from the reality. However, because the origin of the deviation is $\left\langle A^2 \right\rangle_\nu$ as shown above, if the second moment of $\{A_j\}_{j\in\Omega_\nu}$ can be evaluated correctly, the deviation from the reality vanishes. Although the time evolution of the ensemble average of $\{A_j\}_{j\in\Omega_\nu}$, i.e., $A_\nu$, is exactly given by Eq.~(\ref{exact_dyn}), the dynamics for the second moment, i.e., $\left\langle A^2\right\rangle_\nu$, is unknown. In order to obtain $\left\langle A^2\right\rangle_\nu$ without any assumption, an infinitely hierarchical calculation is required. Even if $\left\langle A^2 \right\rangle_\nu$ is exactly given at the initial time, $A_\nu^2$ predicted by Eq.~(\ref{SKOAA}), which plays the role of an approximation of $\left\langle A^2\right\rangle_\nu$ in Eq.~(\ref{exact_dyn}), gradually departs from the true value of $\left\langle A ^2\right\rangle_\nu$. For this reason, the deviation accumulates over time.

\section{Stability of solutions}\label{secStability}
Under the extended OAA, the variable $A_\nu$ obeys the dynamics given by Eq.~(\ref{SKOAA}), where the Kuramoto order parameter $z$ is evaluated using Eq.~(\ref{def_z}).  
The conventional OAA corresponds to the case of $M=1$, and the evolution equation for $z$ is given as 
\begin{eqnarray}
\dot{z} = i\omega z + \dfrac{K}{2}z\left(e^{-i\alpha} - |z|^2 e^{i\alpha}\right)\,.\label{convOAz}
\end{eqnarray}
Because $|z| \le 1$ and the real part of the coefficient for $z$ in the right-hand side of Eq.~(\ref{convOAz}) give the growth rate of $|z|$, $|z|$ increases monotonically to $|z| \to 1$ when the coupling is attractive, i.e., $\cos \alpha > 0$. Therefore, in this case, all the oscillators become in phase. On the other hand, if the coupling is repulsive, i.e., $\cos\alpha < 0 $, $|z|$ decreases monotonically to $|z| \to 0$. In the conventional OAA, because identical oscillators are considered and $z = A_\omega = \exp\left (-\gamma + i \mu\right) $ in this case, $z \to 0$ implies a uniform distribution of oscillator phases.  However, systems with repulsive coupling exhibit complicated behaviors~\cite{hansel1993clustering, golomb1992clustering, gong2019repulsively, abrams2004chimera}, even if the system consists of identical oscillators. A typical example was given by a cluster solution~\cite{golomb1992clustering, gong2019repulsively}. In the repulsive case, there is a possibility that multiple clusters exist and cancel each other's phase effects to satisfy $z = 0$, but this phenomenon cannot be described by the conventional OAA, in which $z = 0$ implies a uniform phase distribution.  

In contrast to the conventional OAA, the extended OAA retains the possibility of a nontrivial phase distribution satisfying $z=0$.  By decomposing $A_\nu$ into $z$ and its variation around $z$ as $A_\nu = z + \Delta_\nu$, the dynamics for $z$ and $\Delta_\nu$ are obtained as 
\begin{eqnarray}
\dot{z} &=& i\omega z + \dfrac{K}{2}z \left(e^{-i\alpha} - |z|^2 e^{i\alpha}\right) \nonumber\\
&-& \dfrac{K}{2} e^{i\alpha} z^\ast\displaystyle\sum_{\sigma=1}^M r_\sigma \Delta_\sigma^2\,,\\
\dot{\Delta}_\nu &=& i\omega\Delta_\nu - K|z|^2 e^{i\alpha}\Delta_\nu \nonumber\\
&+& \dfrac{K}{2}e^{i\alpha} z^\ast\left(\displaystyle\sum_{\sigma=1}^M r_\sigma \Delta_\sigma^2 - \Delta_\nu^2\right)
\end{eqnarray}
with $\sum_{\nu=1}^M r_\nu \Delta_\nu = 0$.  
Note that, in the dynamics of $z$ for the extended OAA, a nonlinear term with respect to $\Delta_\nu$ has been added to the dynamics of $z$ for the conventional OAA~Eq.~(\ref{convOAz}). Because of this nonlinearity, we cannot simply conclude that repulsive coupling, i.e., $\cos\alpha < 0$, leads to a completely desynchronized solution. By taking a complicated initial distribution, a nontrivial behavior that is not predicted by the conventional OAA may occur. 

As in Eq.~(\ref{SKOAA}), different groups interact only through the Kuramoto order parameter $z$. Therefore, if $A_\nu = A_{\nu'}$ ($\nu\neq\nu'$) is realized at a certain moment, the two groups $\Omega_\nu$ and $\Omega_{\nu'}$ exhibit the same time evolution afterwards. Since $A_\nu$ completely determines the phase distribution in $\Omega_\nu$ under the OAA, $A_\nu = A_{\nu'}$ implies no distinction exists between the two groups. Therefore, if $A_\nu = A_{\nu'}$ holds, the two groups can be regarded as united. To keep the two groups distinguishable, $A_\nu- A_{\nu'}\neq 0$ must always hold. Because the evolution of $A_\nu - A_{\nu'}$ is given as 
\begin{eqnarray}
\dot{A}_\nu - \dot{A}_{\nu'} = \left[i\omega - \dfrac{K}{2}e^{i\alpha}z^\ast \left(A_\nu + A_{\nu'}\right)\right](A_\nu - A_{\nu'})\,,\nonumber\\
\end{eqnarray}
the growth rate $\Lambda_{\nu\nu'}$ for $|A_\nu - A_{\nu'}|$ is obtained as 
\begin{eqnarray}
\Lambda_{\nu\nu'} = -\dfrac{K}{2}{\rm Re}\left[e^{i\alpha}z^\ast (A_\nu + A_{\nu'})\right]\,.\label{growth}
\end{eqnarray}
If $\Lambda_{\nu\nu'} <0$, the two groups $\Omega_\nu$ and $\Omega_{\nu'}$ become indistinguishable. 
On the other hand, if $\Lambda_{\nu\nu'} > 0$ is satisfied, $|A_\nu-A_{\nu'}|$ monotonically increases. Because $A_\nu$ has the restriction $|A_\nu|\le 1$, the quantity $|A_\nu-A_{\nu'}|$ has an upper bound. Therefore, $\Lambda_{\nu\nu'} = 0$ is achieved after a long time to stop the growth of $|A_\nu-A_{\nu'}|$ in the case of $\Lambda_{\nu\nu'} > 0$. Generally, $|A_\nu - A_{\nu'}|$ is bounded as $0\le |A_ \nu-A _ {\nu'}|\le 2$, but $\Lambda_{\nu\nu'} = 0$ may be satisfied before $|A_ \nu-A_{\nu'} |$ reaches its upper or lower bound, i.e., 2 or 0. For such solutions, their existence and stability should be investigated in detail.

\subsection{Steady solutions}
In the present paper, a steady solution with constant $|A_\nu|$ for all $\nu$ is considered. Because
\begin{eqnarray}
\dfrac{d}{dt}|A_\nu|^2 = K\left(1 - |A_\nu|^2\right){\rm Re}\left(e^{i\alpha}z^\ast A_\nu\right)\,,\label{Anorm}
\end{eqnarray}
such a solution satisfies $|A_\nu| = 1$ or ${\rm Re}\left(e^{i\alpha}z^\ast A_\nu\right) = 0$.  Thus, the trivial solution is given by $z = 0$. On the other hand, for the solution of ${\rm Re}\left(e^{i\alpha}z^\ast A_\nu\right)=0$ with $z\neq 0$, the angle between $A_\nu$ and $z$ on the complex plane is always fixed as $-\alpha \pm \pi / 2$. However, if such a fixing of phases is realized for all $\nu$, the phase of $z$ is inconsistently given by the linear combination $z = \sum_{\nu=1}^M r_\nu A_\nu$. Thus, in the case of $z\neq 0$, there should be several groups satisfying ${\rm Re}\left(e^{i\alpha}z^\ast A_\nu \right) \neq 0$. Because another choice of a steady $|A_\nu|$ is given by $|A_\nu|=1$, the steady solution with $z\neq 0$ is given by ${\rm Re}\left(e^{i\alpha}z^\ast A_\nu\right) = 0$ for $\nu = 1, 2, \cdots, M_1$ and $|A_\nu|=1$ for $\nu = M_1 + 1, \cdots, M$, where $0 \le M_1 < M$ is a constant integer. Therefore, the steady solution is classified into two types: (i) a trivial stationary solution, i.e., $z = 0$, and (ii) a nontrivial steady solution, i.e., ${\rm Re}\left(e^{i\alpha}z^\ast A_\nu\right) = 0$ for $\nu = 1, \cdots, M_1$ and $|A_\nu| = 1$ for $\nu = M_1 + 1, \cdots, M$ with $z\neq 0$. 

For stable $|A_\nu| = 1$ ($\nu = M_1 + 1, \cdots, M$), when $|A_\nu|$ is initiated as  $|A_\nu| = 1-\delta$ with a small positive parameter $\delta$, $|A_\nu|$ must increase afterwards. Because $K\left(1-|A_\nu|^2\right) > 0$ in such a situation, according to Eq.~(\ref{Anorm}), ${\rm Re}\left(e^{i\alpha}z^\ast A_\nu\right) > 0$ should hold to increase $|A_\nu|$. Note that ${\rm Re}\left(e^{i\alpha}z^\ast A_\nu\right) = 0$ holds for $\nu = 1, \cdots, M_1$ for the nontrivial solution $z\neq 0$. Then, according to Eq.~(\ref{growth}), the growth rate of $|A_\nu - A_{\nu'}|$ is given as $\Lambda_{\nu\nu'} < 0$ for $\nu\in\{1, \cdots, M \}$ and $\nu'\in\{M_1 + 1, \cdots, M \}$.  Thus, the groups in the nontrivial solution merge into one group. The stability of the solution $|A_\nu| =  1$ for all $\nu$ ($ M_1 = 0 $) was discussed previously \cite{gong2019repulsively}. In this case, multiple phase-locked groups merge into one phase-locked group. On the other hand, for the solution with $M_1 \ge 1$, $A_\nu$ for $\nu \in \{1, \cdots, M_1\}$ and $A_{\nu'}$ for $\nu' \in \{M_1 + 1, \cdots, M \}$ must be at an angle different from $z$. However, because $\Lambda_{\nu\nu'} < 0$, $A_\nu$ and $A_{\nu'}$ approach each other. This is an obvious contradiction. Therefore, it is concluded that the solution with $M_1 \ge 1$ cannot exist stably. Only a nontrivial stable solution satisfying $A_1 = \cdots = A_M$ with $|A_\nu| = 1$, i.e., $|z| = 1$, is possible. 

It is concluded that stable steady solutions are classified into two categories: (i) $z = 0$, which includes nontrivial solutions such as cluster and chimera solutions satisfying $A_\nu\neq 0$ as well as the well-known uniform distribution $A_\nu = 0$ for all $\nu$, and (ii) the phase-locked solution $|z|=1$, which can be analyzed by the conventional OAA. Because the stability of $|A_\nu| = 1$ was assumed in the above argument, the stability of each steady solution must be analyzed in detail. With the conventional OAA, for the phase-locked solution $|z| = 1$, it is known that the solution is stable in the attractive case and unstable in the repulsive case. On the other hand, for the solution $z=0$, the stability analysis requires a framework beyond the conventional OAA. The stability to a perturbation for the solution $z=0$ will be discussed in the next subsection. 

Note that the argument on the steady solutions does not depend on the value of $M$. Because the choice $M=N$ reproduces the original $N$-body dynamics (\ref{1body}), only two steady solutions are possible, i.e., $z=0$ or $|z|=1$, even in a system of a finite number of oscillators. 

\subsection{Stability to perturbation}
In order to discuss the stability of a solution after a long time, it is necessary to track the evolution of a small perturbation $\delta A_\nu$ around the solution after a long time $A_\nu^{\rm st}$. $\delta A_\nu$ follows the evolution equation: 
\begin{eqnarray}
\delta \dot{A}_\nu &=& i\omega\delta A_\nu + \dfrac{K}{2} e^{-i\alpha}\delta z-\dfrac{K}{2} e^{i\alpha}\left(A_\nu^{\rm st}\right)^2 \delta z^\ast\nonumber\\
&-&K e^{i\alpha}z^\ast_{\rm st}A_\nu^{\rm st}\delta A_\nu\,,\label{stability}
\end{eqnarray}
where $z_{\rm st} = \sum_{\nu = 1}^M r_\nu A_\nu^{\rm st}$ and $\delta z = \sum_{\nu = 1}^M r_\nu\delta A_\nu$. 
When discussing the stability, it is often useful to denote the order parameters as  $z_{\rm st} = R e^{i\Theta}$ and $A_\nu^{\rm st} = R_\nu e^{i\Theta_\nu}$. However, the phases $\Theta$ and $\Theta_\nu$ are ill-defined when $R = 0$ and $R_\nu = 0$, respectively. When discussing the stability of asynchronous states $R = 0$ or $R_\nu = 0$, such notations of the order parameters cause inconvenience. In general cases, Eq.~(\ref {stability}) is suitable for discussing the stability of solutions. Equation~(\ref{stability}) can be expressed simply using a $2M \times 2M$ matrix $W$ as follows: 
\begin{eqnarray}
\dot{B}_\nu = \displaystyle\sum_{\nu'=1}^{2M} W_{\nu\nu'} B_{\nu'}\,,\label{stability_matrix}
\end{eqnarray}
where $W_{\nu\nu'}$ is given as 
\begin{eqnarray}
&&W_{\nu \nu'} = \left(i\omega - Ke^{i\alpha}z^\ast_{\rm st} A_\nu^{\rm st}\right)\delta_{\nu\nu'} + \dfrac{K}{2}e^{-i\alpha}r_{\nu'}\,,
\\
&&W_{\nu, \nu' + M} = -\dfrac{K}{2}e^{i\alpha}\left(A_\nu^{\rm st}\right)^2 r_{\nu'}\,,\\
&&W_{\nu + M, \nu'} = W_{\nu, \nu' + M}^\ast\,,\\
&&W_{\nu+M, \nu' + M} = W_{\nu, \nu'}^\ast
\end{eqnarray}
for $1\le \nu, \nu' \le M$.  $B_\nu$ is defined as $B_\nu = \delta A_\nu$ for $1 \le \nu \le M$ and $B_\nu = \delta A_{\nu-M} ^\ast$ for $M + 1 \le \nu \le 2M$. 

Now, let us focus on the stability of the solution $z_{\rm st} = 0$. In this case, from Eq.~(\ref{SKOAA}), $A_\nu$ ($\nu = 1, 2, \cdots, M$) rotates at a constant speed on a circle with a constant radius. Therefore, the Lyapunov exponent of the whole system can be easily calculated as the average of the eigenvalues of $W$ on a constant $|A_\nu|$. 
From Eq.~(\ref{stability}), it is found that the perturbation satisfying $r_\nu\delta A_\nu + r_{\nu'}\delta A_{\nu'} = 0$ and $\delta A_{\nu''} =  0$ for all $\nu'' \neq \nu, \nu'$ gives the eigenstate of $W$ with its eigenvalue $i\omega$ because $z_{\rm st}=0$ and $\delta z = 0$. Considering all pairs of $(\nu, \nu')$, such eigenvalues are $(M-1)$-fold. Because $W$ has eigenvalues of complex conjugate pairs, $\pm i\omega$ gives $2(M-1)$ of the $2M$ eigenvalues of the matrix $W$. Because ${\rm tr}W = K \cos\alpha$ when $z_{\rm st} = 0$, the real parts of the remaining two nontrivial eigenvalues are $\frac{K}{2}\cos\alpha$. Thus, the largest Lyapunov exponent is given by $\frac{K}{2}\cos\alpha$ for $\cos\alpha > 0$ and zero for $\cos\alpha < 0$ with $M \ge 2$. Therefore, $z_{\rm st} = 0$ is an unstable solution for the attractive case. Note that this conclusion does not depend on a realization of $\{A_\nu^{\rm st}\}$. Not only a solution with a uniform phase distribution, but also cluster solutions in which $z_{\rm st}=0$ with $A_\nu^{\rm st} \neq 0$ for $^\exists \nu$ are unstable in the attractive case. 
On the other hand, for the repulsive case, $z_{\rm st} = 0$ yields a stable limit cycle for $M \ge 2$. Note that $M=1$ is a special case, where the largest Lyapunov exponent is given by $\frac{K}{2}\cos\alpha$, which implies the uniformly distributed solution is stable in the repulsive case.  Therefore, a nontrivial solution satisfying $z_{\rm st} = 0$ may be realized as a stable limit-cycle solution in the repulsive case. These conclusions in the repulsive case are independent of the realization of $\{A_\nu^{\rm st}\}$. Thus, a cluster solution in which $A_\nu\neq 0$ for all $\nu$ as well as a chimera solution in which $A_\nu\neq 0$ and $A_{\nu'} = 0$ for $^\exists\nu'(\neq\nu)$ can be realized. Recalling that $M = N$ reproduces the dynamics of each oscillator, the above arguments hold even in the case of a finite number of oscillators.  

\section{Numerical results: Cluster and chimera-like solutions}\label{secCluster}
Finally, in order to show the advantage of the higher-dimensional version of the OAA, let us consider a variety of solutions including a cluster solution~\cite{golomb1992clustering, gong2019repulsively} and chimera-like solution~\cite{abrams2004chimera} in the Kuramoto--Sakaguchi model~\cite{sakaguchi1986soluble} of identical phase oscillators.  
In the study of oscillator systems, a cluster and chimera are frequently used terms, and their definitions must be clarified. 
In this paper, a cluster refers to a group of oscillators belonging to a CLD, although it has been defined as a group of oscillators with zero phase difference in many previous reports. In \cite{gong2019repulsively}, Gong et~al. showed that in the Kuramoto--Sakaguchi model consisting of identical oscillators with repulsive coupling, multiple clusters, each of which is characterized by zero phase difference, cannot exist stably. It is worth mentioning that such analyses focused on the oscillators with zero phase difference only. As mentioned in the previous section, the phase-locked solution $|A_\nu|=1$ for all $\nu$ is unstable. However, there remains a possibility where multiple groups with distributed phases exist. 
To handle such situations in this paper, a cluster is defined as a group of oscillators belonging to a CLD. This definition includes the conventional definition of a cluster characterized by zero phase difference. On the other hand, the chimera state was defined in \cite{abrams2004chimera} as ``an array of identical oscillators splits into two domains: one coherent and phase locked, the other incoherent and desynchronized.'' However, from the standpoint mentioned above, the ``phase locked'' condition is too strict to characterize a coherent state. In a subdomain of various systems,  a coherent state would be realized with a local Kuramoto order parameter $0 < |A_\nu| \le 1$. Thus, in this paper, by using our definition of clusters, a chimera state is defined as a state in which the clusters of $A_\nu \neq 0$ and $A_\nu = 0$ coexist. Below, the coexistence of clusters with $A_\nu \neq 0$ and $A_\nu \simeq 0$ is numerically shown. However, such a state is called a chimera-like state in this paper because the exact asynchrony $A_\nu = 0$ cannot be confirmed numerically.

\begin{figure}
\centering
\includegraphics[width=8.6cm]{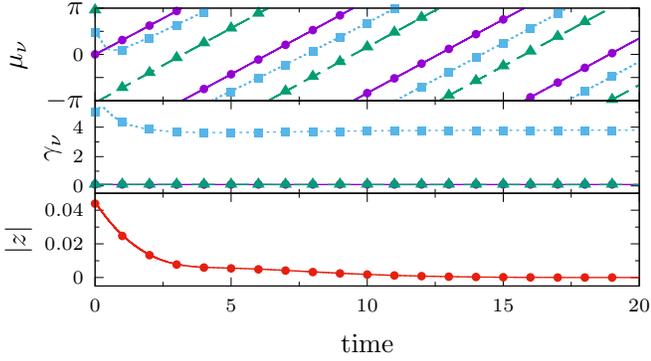}
\caption{(Color online) Chimera-like cluster solution of three groups of oscillators with the same natural frequency $\omega$ in the repulsive coupling case. Top panel: time evolution of $\mu_\nu$. Middle panel:  time evolution of $\gamma_\nu$. Bottom panel: time evolution of Kuramoto order parameter $z$. The parameters are set to $\omega=1.0, K=1.0, \alpha=3\pi/4,$ and $r_1 = r_2 = r_3 = 1/3$. The initial values are set to $\gamma_1 = \gamma_2 = 0.1, \gamma_3 = 5.0, \mu_1 = 0.0, \mu_2 = 3.0,$ and $\mu_3 = 1.5$.  The solid, dashed, and dotted lines in the top and middle panels correspond to the solutions of Eq.~(\ref{SKOAA}) for $A_1$, $A_2$, and $A_3$, respectively.  The circle, triangle, and square marks in the top and middle panels correspond to the numerical solutions of Eq.~(\ref{original_dyn}) for $A_1$, $A_2$, and $A_3$, respectively. In the bottom panel, the solid line and circle marks correspond to the solutions of Eq.~(\ref{SKOAA}) and Eq.~(\ref{original_dyn}), respectively.   In the calculation of Eq.~(\ref{original_dyn}), the numbers of oscillators were taken as $N_1 = N_2 = N_3 = 10^5$.  \label{fig_M3cluster}}
\end{figure}

\begin{figure}
\centering
\includegraphics[width=8.6cm]{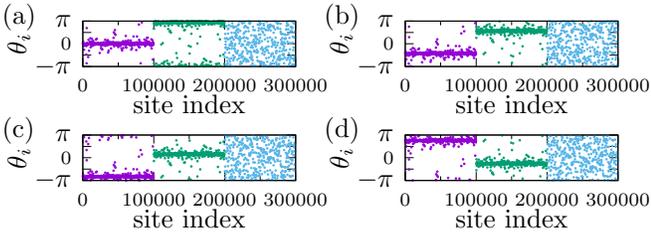}
\caption{Phase distributions obtained from the direct calculation of Eq.~(\ref{original_dyn}) corresponding to the result shown in Fig.~\ref{fig_M3cluster}.  The site indices $0$ to $10^5-1$, $10^5$ to $2\times 10^5 -1$, and $2\times 10^5$ to $3\times 10^5 -1$ correspond to the oscillators belonging to the clusters of $A_1$, $A_2$, and $A_3$, respectively.  The panels (a)--(d) correspond to the snapshots at $t=0, 5, 10, 15$, respectively.  For the convenience of viewing, the points are plotted by thinning out at a rate of $1/250$.  \label{fig_M3Raster}}
\end{figure}

In the case of $M = 3$, a nontrivial cluster solution with $z=0$ is numerically observed.  The solution of Eq.~(\ref{SKOAA}) and the direct numerical solution of Eq.~(\ref{original_dyn}) for the repulsive case are shown in Fig.~\ref{fig_M3cluster}. The solutions for Eqs.~(\ref{SKOAA}) and (\ref{original_dyn}) agree very well.  For the solutions of Eq.~(\ref{SKOAA}), $\mu_\nu$ and $\gamma_\nu$ are evaluated as $\mu_\nu = \arg A_\nu$ and $\gamma_\nu = -\ln|A_\nu|$, respectively. 
Further, the snapshots of the corresponding oscillator phases obtained using Eq.~(\ref{original_dyn}) are shown in Fig.~\ref{fig_M3Raster}.  The oscillators in the two clusters corresponding to $A_1$ and $A_2$ are almost anti-phase, and the oscillators in the cluster corresponding to $A_3$ are almost uniformly distributed after a long time.  Note that such a chimera-like cluster solution cannot be obtained by the conventional OAA as mentioned in the previous section.  
Note also that the numerical solution of Eq.~(\ref{original_dyn}) with finite $N$ and that of Eq.~(\ref {SKOAA}) obtained in the thermodynamic limit $N\to\infty$ are stable, as mentioned in the previous section.
When the initial values for $A_\nu$ ($\nu = 1, 2, 3$) are identical, the results are the same as in the conventional OAA, i.e., $A_1$, $A_2$, and $A_3$ all correspond to characteristic functions of uniform distributions; in other words, $A_\nu = 0$. As the difference between the initial values of $A_\nu$ increases, the solutions $A_1$, $A_2$, and $A_3$ gradually split, resulting in a nontrivial cluster solution.  

\begin{figure}
\centering
\includegraphics[width=8.6cm]{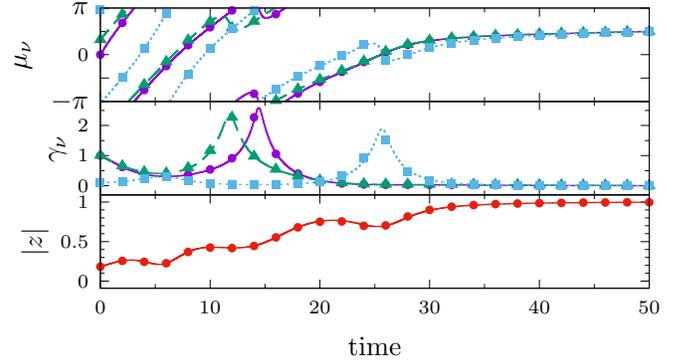}
\caption{(Color online) Nontrivial evolution of three groups of oscillators with the same natural frequency $\omega$ in the attractive coupling case. The parameters are set to $\omega=1.0, K=1.0, \alpha=1.8\pi/4,$ and $r_1 = r_2 = r_3 = 1/3$. The initial values are set to $\gamma_1 = \gamma_2 = 1.0, \gamma_3 = 0.1, \mu_1 = 0.0, \mu_2 = 1.0,$ and $\mu_3 = 3.0$.  \label{fig_M3attractive}}
\end{figure}

\begin{figure}
\centering
\includegraphics[width=8.6cm]{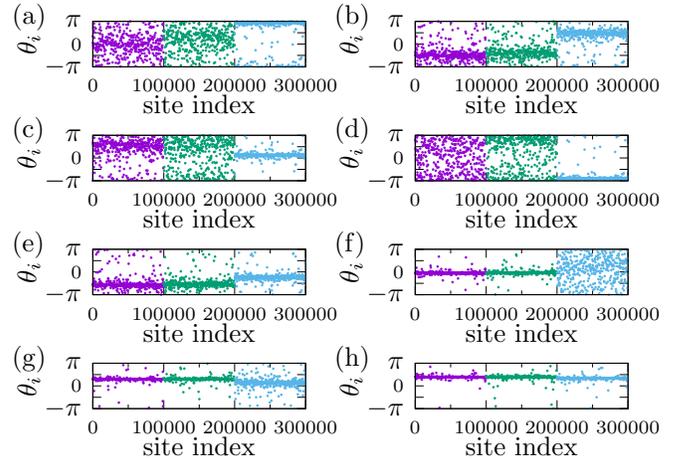}
\caption{Phase distributions obtained from the direct calculation of Eq.~(\ref{original_dyn}) corresponding to the result shown in Fig.~\ref{fig_M3attractive}.  The panels (a)--(h) correspond to the snapshots at $t=0, 5, 10, 15, 20, 25, 30, 35$, respectively.  \label{fig_M3attractiveRaster}}
\end{figure}

Nontrivial dynamical behavior is observed even in the attractive coupling case, as shown in Fig.~\ref{fig_M3attractive}.  The Kuramoto order parameter $|z|$ increases with oscillation, whereas $|z|$ increases monotonically in the prediction of the conventional OAA.  As shown in Fig.~\ref{fig_M3attractiveRaster}, it is possible to predict the chimera-like state in the transient regime by using the extended version of the OAA.  As mentioned in the previous section, multi-cluster or chimera-like solutions with $z=0$ are forbidden as stable solutions in the attractive case. 

Several numerical experiments have shown that the stability of $z$ is independent of $M$. Synchronous solutions of $|z| = 1$ were obtained for the attractive case, i.e., $\cos\alpha > 0$, and asynchronous solutions of $z=0$ were obtained for the repulsive case, i.e., $\cos\alpha < 0$, for several initial conditions.  In other words, the behavior of the Kuramoto order parameter $z$ after a long time is independent of the initial condition.  Thus, the stability of the solutions discussed in section~\ref{secStability} has been validated numerically.  

\section{Conclusion}\label{sec_conclusion}
To conclude, the OAA, which is a method for reducing the high number of degrees of freedom of globally coupled phase oscillators to a two-dimensional manifold, has been extended for reduction to a high-dimensional manifold.  The conventional two-dimensional Ott--Antonsen manifold has been clarified to be a manifold of a CLD, which is characterized by two parameters.  Owing to the representative property of the poles of a CLD, the many-body problem of phase oscillators has been reduced to a single oscillator problem under the conventional OAA.  By taking advantage of the linearity of the characteristic function with respect to the superposition of empirical distributions, the extension of OAA has been realized by the superposition of CLDs.  
Since the extended OAA is exact in the thermodynamic limit, it would be a powerful tool to investigate the behaviors of a system consisting of many phase oscillators.  
Moreover, this extension enables the systematic approximation of the behavior of coupled phase oscillator systems with arbitrary initial conditions.  

The extended OAA has been employed for the Kuramoto--Sakaguchi model of identical phase oscillators to show a variety of dynamical behaviors. It has been shown that cluster and chimera states, which cannot be obtained by the conventional OAA, exist in the Kuramoto--Sakaguchi model in the repulsive regime. From a linear stability analysis, these states were found to be stable. The conventional chimera state was found in systems with couplings dependent on the distances between the oscillators~\cite{abrams2004chimera}. It was previously believed that the chimera state could stably exist only in the presence of intermediate nonlocal couplings and neither global nor local couplings. However, it has been shown that chimera states can also exist stably in the presence of all-to-all couplings, where there is no concept of distance. This fact may deepen the understanding of the origin of complicated behaviors of oscillator systems.  

The proposed method can be applied to a wide range of phase oscillator systems. In this paper, systems consisting of identical oscillators without noise have been analyzed. However, our proposed method can be applied to non-identical cases, where the natural frequencies are distributed, as well as to systems under the influence of common noise. Further, it can also be applied to systems with a time delay by using an approach similar to that shown by Ott and Antonsen~\cite{ott2009long}. The proposed method is applicable to all systems where the conventional OAA can be applied. Since our method restricts the phase distribution of the system to a superposition of CLDs, it is not exact in the presence of noncommon noise. Further studies are required to apply the proposed method to such a noisy case. 

Although the stability of cluster and chimera states has been discussed in this paper, the question of which cluster or chimera state will appear spontaneously as a result of relaxation remains to be solved. In addition, this question is related to the natural number of clusters.  In our analysis, the number of clusters $M$ was given. However, in general, $A_\nu = A_{\nu'} $ can be realized, and the number of clusters can change. It is necessary to discuss the stability of the solution $A_\nu - A_{\nu'}$ in detail. The investigations for natural nontrivial solutions and numbers of clusters are topics for future studies. 

\begin{acknowledgments}
This work was supported by JSPS KAKENHI grants numbered JP17H06469 and JP19K20360. 
\end{acknowledgments}


\end{document}